\documentclass[eprint]{actapoly}

\usepackage{graphicx}
\usepackage{amsmath}
\usepackage{mathrsfs}
\usepackage{hyperref}

\begin{document}

\title[\v{Z}ampa's systems theory]
{\v{Z}ampa's systems theory: a comprehensive theory of measurement in dynamic systems}

\correspondingauthor[R. Rycht\'{a}rikov\'{a}]{Renata Rycht\'{a}rikov\'{a}}{my}{rrychtarikova@frov.jcu.cz}
\author[J. Urban]{Jan Urban}{my}
\author[D. \v{S}tys]{Dalibor \v{S}tys}{my}

\institution{my}{University of South Bohemia in \v{C}esk\'{e} Bud\v{e}jovice, Faculty of Fisheries and Protection of Waters, South Bohemian Research Center of Aquaculture and Biodiversity of Hydrocenoses, Kompetenzzentrum MechanoBiologie in Regenerativer Medizin, Institute of Complex Systems, Z\'{a}mek 136, 373 33 Nov\'{e} Hrady, Czech Republic}

\begin{abstract}
The article outlines in memoriam Prof. Pavel \v{Z}ampa's concepts of system theory which enable us to devise a measurement in dynamic systems independently of the particular system behaviour. From the point of view of \v{Z}ampa's theory, terms like system time, system attributes, system link, system element, input, output, sub-systems, and state variables are defined. In Conclusions, \v{Z}ampa's theory is discussed together with another mathematical approaches of qualitative dynamics known since the 19$^{\mbox{th}}$ century. In Appendices, we present applications of \v{Z}ampa's technical approach to measurement of complex dynamical (chemical and biological) systems at the Institute of Complex Systems, University of South Bohemia in \v{C}esk\'{e} Bud\v{e}jovice.
\end{abstract}


\maketitle

\section{Introduction}
\label{intro}

Human thinking has been evolving to be constructive. It creates models of the Nature perceived by senses, directly or indirectly, and communicates them~\cite{Buquoy1820ideele}. The overall spread of technical products makes them a kind of language which is much more general than any of the national languages. The only competitor of technology in the field of general languages is mathematics which has the advantage of not being connected to a particular technical solution. 

For an overwhelming part of humans, the mathematical notation is difficult for comprehension which leads to the preference of representation of mathematical results by the approximation of formulae to functions of technical tools, in addition, in the current computer age, in the form of computer visualizations. This leads to identification of these technical constructs with mathematical expressions. Such a way of thinking has been more widespread in theoretical and experimental physics than in technology, chemistry or other sciences which are closer to reality. In latter sciences, the mathematical formulae are rather useful constructive tools whose outputs need to be further verified experimentally.

No mathematical physicists of the late 19$^{\mbox{th}}$ century understood their equations as a model of any ultimate truth. They developed mathematical tools as mutually consistent descriptions. In the beginning of the 19$^{\mbox{th}}$ century, there were many thinkers who criticized concrete mechanistic approximations and suggested that theories should be purely mathematically deduced~\cite{Buquoy1819}. The nicest expression of this approach can be found in Poincar\'{e}'s Science and Method~\cite{Poincare1908} where is written "...it is economy of thought that we should aim at, and therefore it is not sufficient to give models to be copied." In other words, in the Nature, there seem to be general rules, which are more constrained than the intellectual freedom of mathematicians, and mathematics is a way how to describe them.

The observation of qualitatively dynamic rules in the Nature became known by the advent of Poincar\'{e}'s qualitative dynamics in the late 19$^{\mbox{th}}$ century~\cite{Poincare1880,chaosbook}. It showed that non-linear dynamic systems can be highly unpredictable so that, upon small changes in starting conditions, their behaviour can be deviated wildly. However, in contrast, a phase space of a dynamic system is always segregated into zones of attraction~\cite{chaosbook} within which the system travels into a limit set -- a region of space within which it stays forever. The famous Poincar\'{e}-Bendixson Theorem applied to a two-dimensional space~\cite{Poincare1892,Bendixson1901} states: "In a differentiable real dynamic system defined on an open subset of a (two-dimensional) plane, every non-empty compact $\omega$-limit set of an orbit, which contains only finitely many fixed points is either a fixed point, a periodic orbit, or a connected set composed of a finite number of fixed points together with homoclinic and heteroclinic orbits connecting these." At higher dimensions, you can have other types of behaviour. For instance, chaotic behaviour can only arise in continuous dynamic systems whose phase spaces have three or more dimensions. In other words, majority of the dynamic objects do not die but, for infinite amount of time, occupy a certain volume of space. Understanding of the Nature can mean a classification of dynamic systems. For this classification, we must understand the process of measurement.

Apart from outlining the work of Prof. Pavel \v{Z}ampa\footnote{Pavel \v{Z}ampa (1936--2006), a former head of the Department of Cybernetics at the Faculty of Applied Science, University of West Bohemia in Pilsen, spent his life in developing technical cybernetics. He graduated in 1962, gained his PhD in 1973, became an assistant professor in 1990 and a full professor in 1997. Since 1990, his publications in English are found to be only sparse \cite{Zampa2004}. The most comprehensive source of his ideas is the revised version of his habilitation which he adopted for his co-workers and students in late 1990'~\cite{ZampaHabilitation}.}, this article aims to demonstrate that it is still, or perhaps increasingly, possible to devise a measurement in dynamic system with a full mathematical rigour. \v{Z}ampa devised a rigorous concept of measurement which is independent of the particular system behaviour, including any observer, and explores limits of experimental cognition.

\section{The system}
\label{system}

\v{Z}ampa concluded that the systems theory is not stabilized in full exactness\footnote{In the text, direct translations of \v{Z}ampa\'{}s work are highlighted by italics.}. \emph{Some authors understand it as a theory of real systems}~\cite{Bertalanffy1950} \emph{while others understand it more as a theory of abstract models of real systems}~\cite{Kalmanetal1969}. ...\emph{Some of the theories are too narrow and do not include all occurring examples}~\cite{Kalmanetal1969,Kalman1960,More1956}, \emph{while others are too broad and in principle inconsistent}~\cite{Mealy1955,ZadehDesoer1963,Zadeh1964} \emph{and must be corrected in an appropriate manner}. ...\emph{In most cases, there are subjectively motivated mathematical requirements which can reflect real needs only sparsely}~\cite{Willems1991}. \emph{This situation requires a urgent solution, mainly in case of continuous systems, in particular the stochastic ones.}

\v{Z}ampa constructs his systems theory by definition of a class of abstract systems which model all real systems. An abstract system exists for each real system.

Next, \v{Z}ampa addresses the question how such a general adequate system should look like. He concludes that \emph{a system, demarcated on the basis of the given paradigm, has no input which could influence the system in any way.} It says that an input and its properties are part of the system. Instead, as shown later, the system can be split into sub-systems which are connected by bonds. The latter approach enables the system to be represented by measurable quantities at any time. 

\emph{Certain substantialities, which are perceived via perceptions and caused by outer environment, are considered as natural phenomena. They demonstrate certain qualities of properties of this environment which is represented by certain objects. Such a quality is, e.g., temperature, location or speed of a certain object. We do not perceive only one perception but their ensembles which we organize in our minds in a certain way and between which we find certain relations. It is rather a certain construct in our mind, which attributes our concepts about the organization of the world to the real world, than a perception of real relations between objects.}

\emph{While the natural phenomena can be, to a certain extent, perceived immediately, their mutual relations cannot. Therefore, a recognition of these relations is much more complicated than the recognition of natural phenomena themselves. In this way, the definition of real system depends on our reasoning.}

\emph{In general, we assume that a set of all events contains primary events, which we immediately consider, and secondary events, which are all other events by which the primary events are affected. It depends on our intuition and, in general, on the correctness of our concepts about the system which events are included into the system. In any case, it is a matter of compromise between the required precision and the complexity of the model. In fact, it is always an incomplete set. However, the only way how to demarcate the system is via the approach of trials and errors which is stopped upon achievement of a certain precision. In this way, we obtain a set of events which we further consider as independent of all other events. This does not mean that events in the system could not affect events outside the system.  Such an affection should be only unidirectional and should not affect the behaviour of the system in any way.}

In addition, \v{Z}ampa's systems theory is an examination of the role of measurement in the system analysis. Moreover, it offers a methodological approach to the mathematical description of cognition. This would like to be demonstrated on the selection of items from Zampa's work.

\subsection{Abstract system}
\label{abstractSystem}

\subsubsection{System time and attributes}

\emph{System events occur at certain objects of the real world and represent themselves by changes of certain qualities which we are able to perceive in a certain way and organize in our mind. A special quality which enables us to organize these events is time.} 

\begin{figure}[!t]
\centering
\includegraphics[width=.8\linewidth]{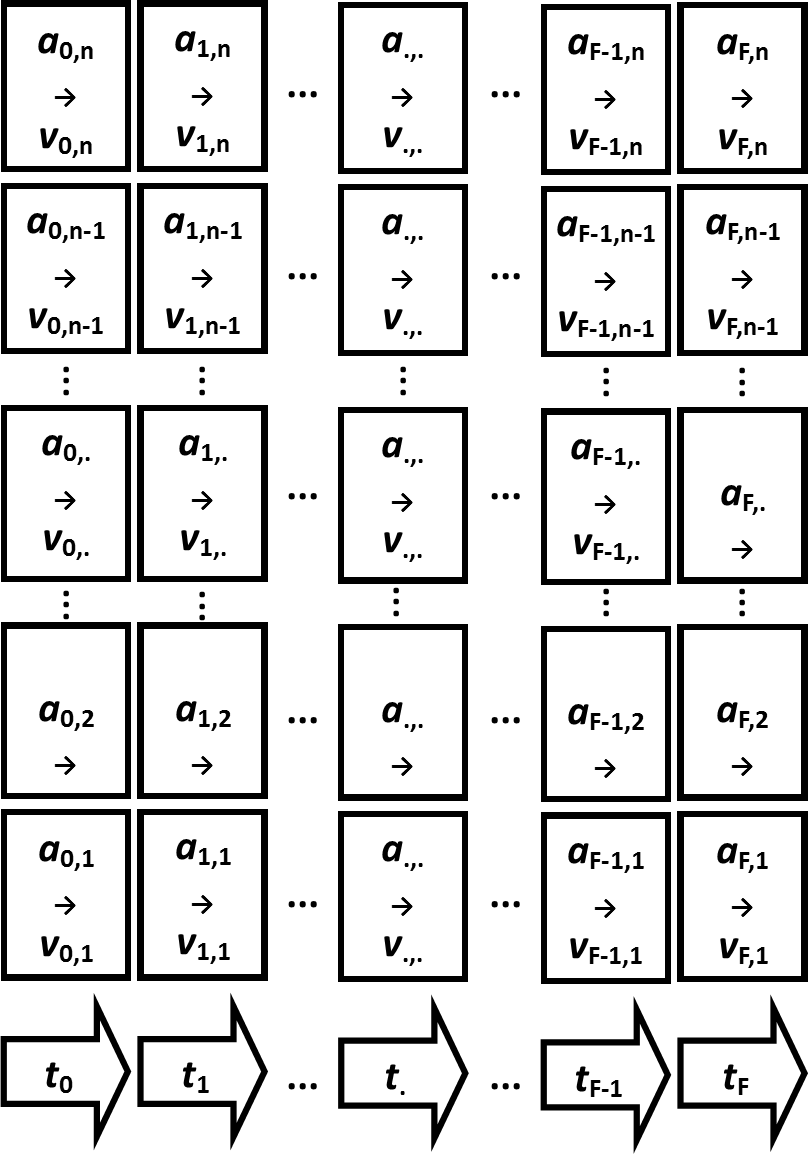}
\caption{The scheme of the concept of the system time, system attributes, and variables. The system time $T$ is an ordered set of time instants $t_i$, $i \in \{0,1,2, ..., F\}$, in which a sharp order exists. The system time can be also understood as an order in the set of system attributes and variables. The system attribute $a_{i,j}$, $i \in \{0,1,2, ..., F\}$, $j \in \{1,2, ..., n\}$, is a certain property of the system, e.g., the position in space, velocity, acceleration. The system variable $v_{i,j}$, $i \in \{0,1,2, ..., F\}$, $j \in \{1,2, ..., n\}$, is a value of scalar or vector character (depicted) by which the system attribute is characterized, i.e., the value of $x$, $y$, and $z$ position, velocity vector, etc.}
\label{fig:System_time_attributes_and_variables}
\end{figure}

Neither us humans, nor any measuring instrument is able to record external quantities in infinitely many time intervals. An adequate model of the perceived time is a real time $t$ whose definition set

\begin{equation}
t \in T
\end{equation}
is a non-empty set $T$ of all time events with $K$ as an appropriate index set (Figure~\ref{fig:System_time_attributes_and_variables}). If we stay in the realm of experimentally verifiable realities, we assume only a limited set of experimental times from $0$ to $F$, i.e.

\begin{equation}
K = \{0,1,2,...F\}.
\end{equation}
In the same time, we require that a complete sharp order

\begin{equation}
t_0 < t_1 < t_2 < ... < t_F,
\end{equation}
which we shall interpret as a precedence of one time instant before the other, is defined at the set $T$.

It should be emphasized that it is not necessary to interpret the set $T$ as the set of time instants. Sometimes it is more appropriate to assume it as a set which determines the order of events in a given system. A special case can be a set which contains only one element. It is important to realize that the system is defined only at the set $T$. Outside this set, the system does not exist for us. 

An abstract system, a good model of a real system, has to include models of all its attributes. Abstract attributes will be denominated by symbols

\begin{equation}
a_i,\mbox{ where }i=1,2,...n, i \in I
\end{equation}
and denote names or designations of studied attributes such as coordinates of position, coordinate of speed, position of a switch, and verity of a statement. A set of all abstract attributes is designated by symbol $A$. An adequate model of such a $i$-th attribute $a_i$ is an \texttt{abstract variable} of the $i$-th attribute

\begin{equation}
v_i\in V_i,\mbox{ where }i\in I,
\end{equation}
whose definition set is a non-empty set $V_i$ with elements called \texttt{values of the i-th attribute}.

\v{Z}ampa's repeated emphasis on the distinction between the system, the measurable variables and model attributes is the key to understanding the contemporarity of \v{Z}ampa's thoughts. The fact that \v{Z}ampa introduced the mathematical formalism into these general philosophical terms enabled him to formalize the whole theory and measurement of dynamic systems. 

In Appendix A,~Figure~\ref{fig:PDGE_series_BZ_model} shows an example of a system's simulation where we know the elementary time step. The trajectory which is in detail reported in~\cite{Stysetal2015b,Stysetal2016b} is by statistical analysis segmented into regions of oscillations between two clusters which can be considered as natural sections of the system's evolution. The time decimation of the system trajectory influenced the result of statistical analysis. This is due to a dominating low-frequency oscillation which affects the value of the variable point divergence gain. By the decimation of the series, i.e., of the evolving system, we loose information about this low-frequency component. The time decimation probably disables to infer the correct model of the system from the given dataset.

\subsubsection{System trajectory}

\begin{figure}[!t]
\centering
\includegraphics[width=.8\linewidth]{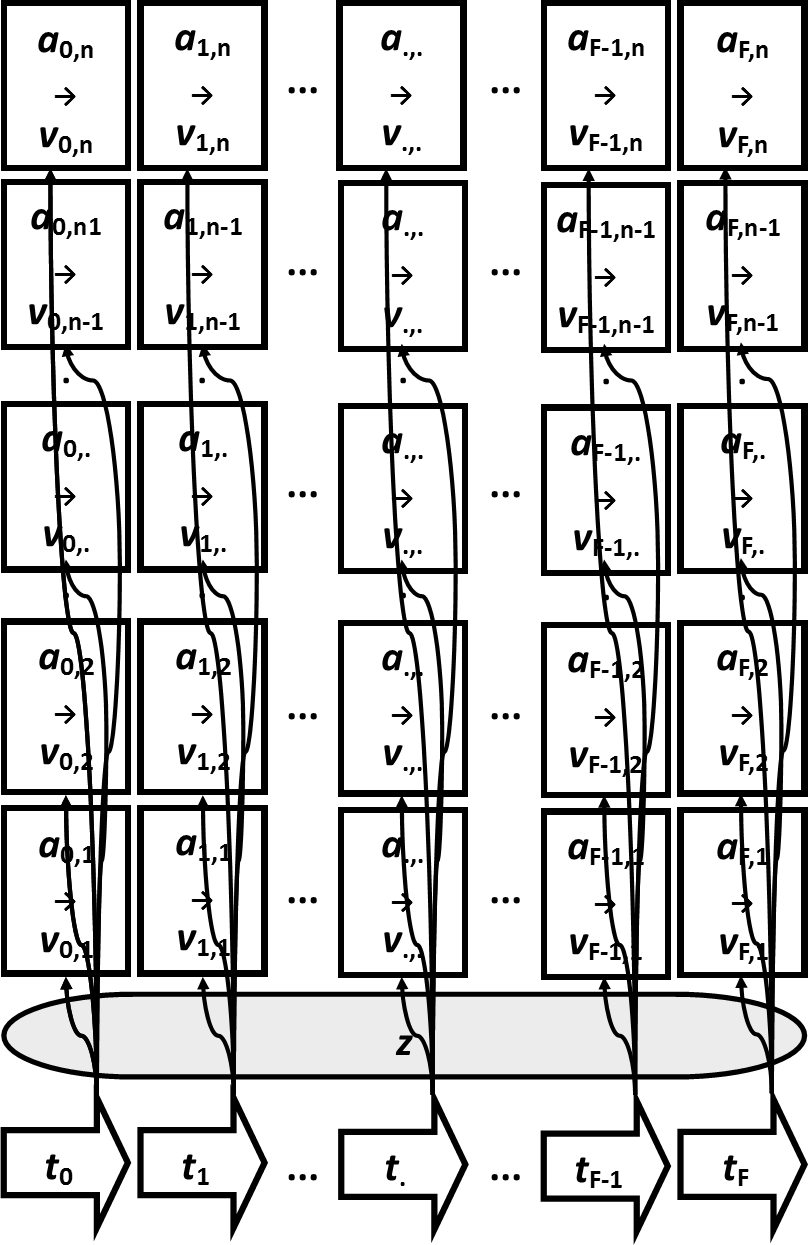}
\caption{The scheme of a system trajectory $z$ as a mapping from the set of time instants $t_i$, $i \in \{0,1,2, ..., F\}$ to the set of attributes $a_{i,j}$, $i \in \{0,1,2, ..., F\}$, $j \in \{1,2, ..., n\}$.}
\label{fig:System_trajectory}
\end{figure}

The obvious property of the dynamic system is its trajectory. To define it, we shall first introduce the term \texttt{system variable} $v$

\begin{equation}
v = (v_1, v_2,..., v_n),
\end{equation}
which is an ordered set of $n$ variables of system attributes. Its definition set

\begin{equation}
v\in V, \mbox{ where } V = V_1 \times V_2 \times... \times V_n.
\end{equation}
An ordered tuple (T,V) is then the basis of the mathematization of the definition of the system trajectory. In order to define the term \texttt{system trajectory}, we must state that, at each system time instant
$t\in T$, each attribute $a_i, i\in I$ of a real system acquires exactly one value $v_i\in V_i$. A system trajectory (Figure~\ref{fig:System_trajectory}) is then a mapping $z$

\begin{equation}
z: T\times I \rightarrow \bigcup_{i\in I} V_i \mbox{ such that }z(t,i)\in V_i,\mbox{ where }i\in I,
\label{eq10}
\end{equation}
from the set $T$ into the sets $A$ and $V$. 

The mapping $z$ is not generally unique. The \texttt{set of all system trajectories} will be designated $\Omega$. If the trajectory realized by the system falls into a given sub-set $B$ of the set $\Omega$, i.e., $B \subset \Omega$, we shall say that an event $B$ occurred on the system. In all other cases, the event $B$ did not happen. 

\begin{figure}[!t]
\centering
\includegraphics[width=\linewidth]{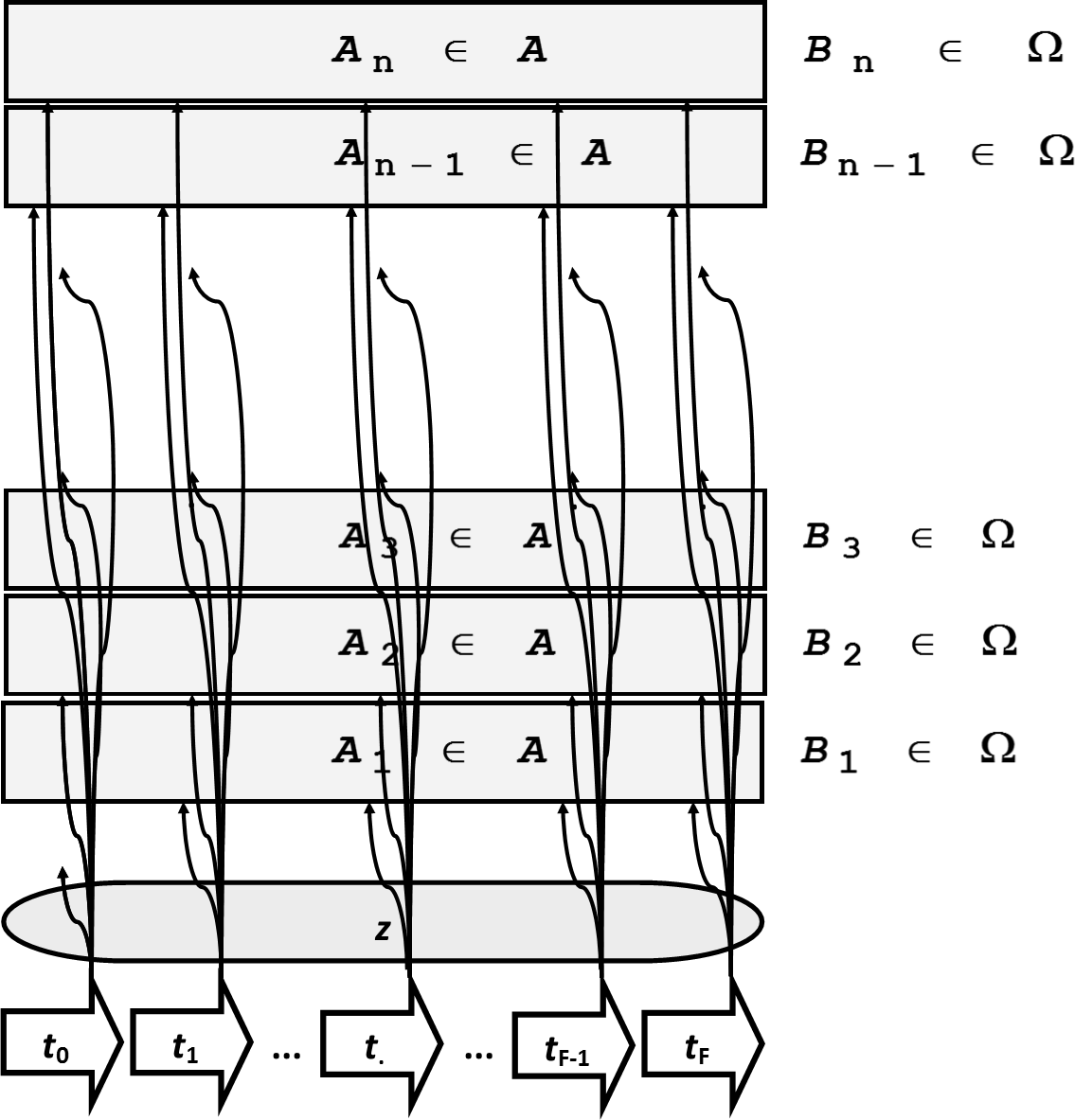}
\caption{The scheme of a system event as a sub-set $B_i$, $i \in \{0,1,2, ..., n\}$ of all trajectories $\Omega$ which satisfy a certain condition. For a selected sub-set of attributes $A_i \in A$, it can acquire a certain set of values of variables.}
\label{fig:System_events}
\end{figure}

The system event is usually a set of trajectories $z$ which have a certain property $V(z)$. For instance, if at time $t_k$ the trajectory of the attribute $a_i$ passes a point $v_i = z(t_k,i)$, we can say that the event

\begin{equation}
B = \{z\mid  z(t_k,i) = v_i\}
\end{equation}
happened. The set of all sub-sets $B$ of the set $\Omega$ is labelled by the symbol $\cal{B}$, where $\cal{B} = \{B|B \subset$ $\Omega$$\}$, and is called \texttt{a set of all system events}.

\section{General abstract system}

\subsection{Definition}

The general abstract system is an ensemble of all trajectories which are outcomes of the mapping $z$. If a set $\cal{B}$ is given, all its elements as well as the set $\Omega$ of all mappings $z$ are also given. Then, the definition set and the set of values of this mapping and thus also sets $T$, $I$, $V_i, i\in I$, and $V$ are also demarcated. On the contrary, by these sets, i.e., by the sets $T$ and $V$, it is possible to define the set $\cal{B}$ and thus the whole abstract system. The abstract system will be further denoted $\mathscr{S}$. The system is unambiguously defined by a tuple of $T$ and $V$, i.e.,

\begin{equation}
\mathscr{S} = (T,V).
\end{equation}

\subsection{System behaviour}

\emph{In the previous chapter, system behaviour which, besides other aspects, describes also qualitative relations between events was defined as the general abstract system ${\mathscr{S}}=(T,V)$... it is useful to complement our knowledge by causal relations between events to determine, e.g., for two selected events, how the first one affects the second one and vice versa or whether their mutual affections are neutral.}

\emph{Dependencies of this kind will be called qualitative dependencies}...

...\emph{The qualitative relations are very important for the definition of the orientation of the cause between system events. They will lead us to the demarcation of structural terms in the systems theory such as \texttt{system link, sub-system, system element, input, output, state variables of the system and system structure}. Later, it will enable us to explain certain experimental experience which was so far explained by a mythical rather than scientific argumentation}... 

It should be noted that the term qualitative dynamics was coined by Poincar\'{e} about 130 years ago~\cite{Poincare1880,chaosbook}. It deals with the qualitative differences in the behaviour of non-linear systems, in Poincar\'{e}'s times mainly continuous. But it was Poincar\'{e} himself who showed how, using so-called Poincar\'{e} sections, the continuous systems can be turned into discrete deterministic and discrete stochastic systems~\cite{chaosbook}. \v{Z}ampa had obviously limited knowledge of these findings, if any. He came up with an idea of qualitative relations in dynamic systems and held it from a quite different point of view.

Poincar\'{e} concludes that we are considerably restricted in the scope of \texttt{general behaviour} of dynamic systems. \v{Z}ampa describes how systems can be \texttt{constructed} in order to be confronted with a perceivable system. Both describe principal mathematical reasons for such restrictions. The current research on dynamics of discrete systems indicates that the scope of available models can be even more constrained~\cite{Wuensche2011,Stysetal2015b} and it is sensible to expect an observation of only a small set of state space trajectories. 

\subsection{Deterministic systems}

An ordered triple

\begin{equation}
\mathscr{D} = (T,V,z)
\end{equation} 
is called \texttt{a deterministic abstract system}.

\subsection{Stochastic systems}

If we define $P(B)$, $B\in\cal{B}$, as a probability of each event $B$ in the set $\cal{B}$ of all sets of events, then the ordered set

\begin{equation}
\mathscr{P} = (T,V,P)
\end{equation}
is called \texttt{a stochastic (abstract) system}.

\subsection{Causal system}

\begin{figure}[!t]
\centering
\includegraphics[width=\linewidth]{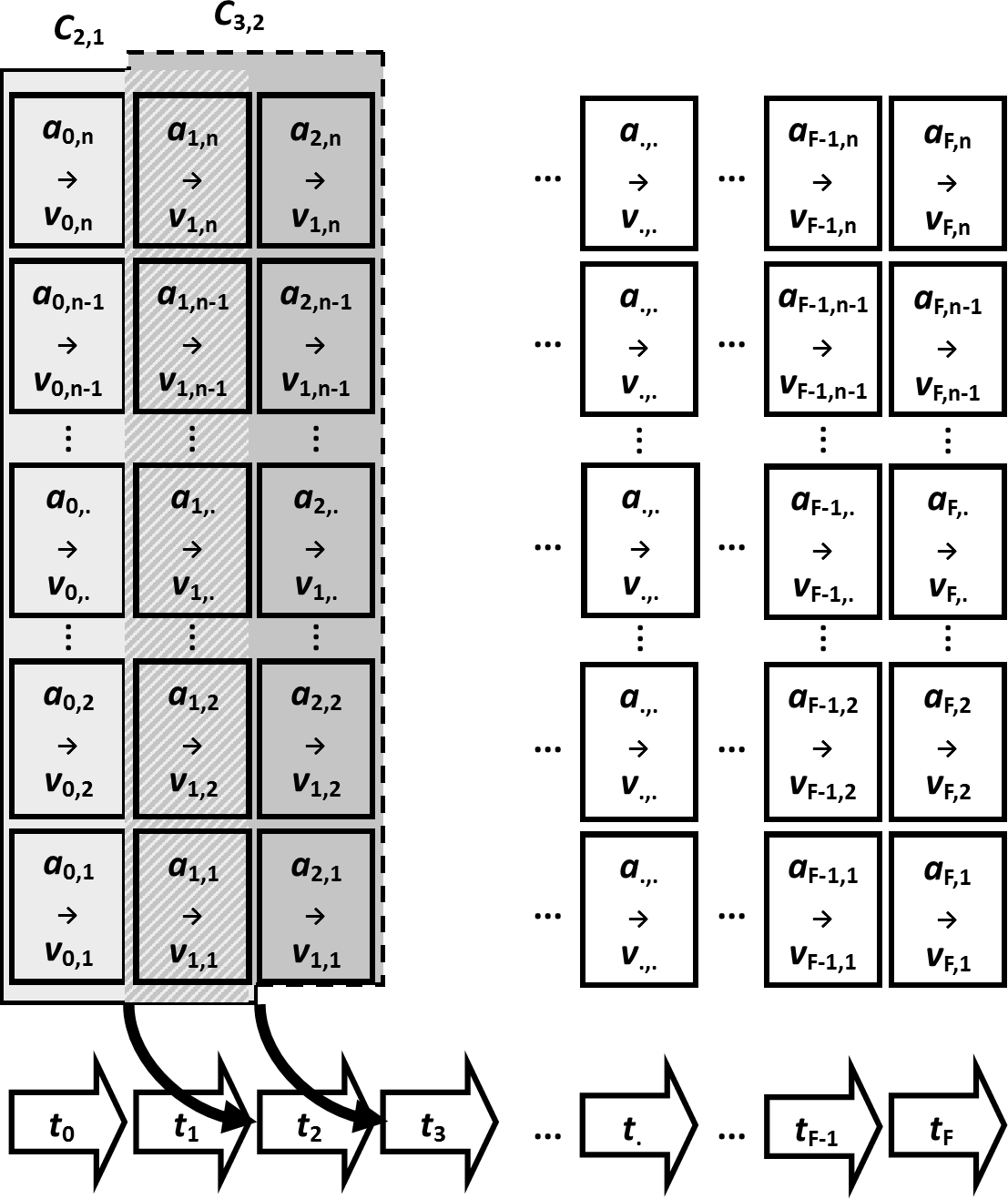}
\caption{The general concept of a causal system. The system mapping \textit{z} determines an order of evolution of state variables $v_{i,j}$, $i \in \{0, 1, 2, ..., F\}$, $j \in \{1, 2, ..., n\}$ in time. $C_{k,l}$ is a complete immediate cause which determines the consequence $D_{k,l}$, $k \in \{2, 3, ..., F\}$, $l \in \{1, 2, ..., F-1\}$.}
\label{fig:Causal_relations}
\end{figure}

We generally assume that the system trajectory is defined as a  mapping $z$ with a definition set

\begin{equation}
D = T \times I
\end{equation}
which is by its internal mechanism defined in parts. We assume that a set

\begin{equation}
{\cal{I}} = \{I_1, I_2, ..., I_m\},
\end{equation}
is a decomposition of the set $I$ at which is defined a convenient order $<$ by the order of determination. Thus holds

\begin{equation}
I = \bigcup_{i = 1}^{m} I_i,\mbox{ where } I_i,I_j \in {\cal{I}}, \ I_i \neq I_j \Rightarrow I_i \cap I_j = \emptyset.
\end{equation}
This aspect is illustrated in Figure \ref{fig:Causal_relations}.

We saw that the trajectory was determined sequentially on different sub-sets of the set $D$ which forms a certain ordered decomposition $\cal{D}$ of the set $D$. It is enough to define each segment $z \mid D_{k,l}$ of the trajectory $z$ exactly once dependently on the segment $z \mid C_{k,l}$, where

\begin{equation}
C_{k,l} \subset \bigcup_{(i,j)<(k,l)} D_{i,j}
\end{equation}
holds. This requirement can be interpreted mainly as a certain natural property according to which the trajectory segment $z \mid D_{k,l}$ is determined by only preceding and not future segments. This is in agreement with the \texttt{causality principle}, where each event has its own cause, which precedes its consequence. Further, we assume that the cause $C_{k,l}$ determines the consequence $D_{k,l}$ completely and is understood as a \texttt{complete immediate cause} of the consequence $D_{k,l}$.

In other words, \v{Z}ampa's definition is the first complete mathematical -- qualitative -- definition of a causal abstract system. It opens the possibility to fully and exactly discuss the real causal system for which the causal system is an adequate \texttt{constructive} model. In the text below, we shall in full examine the definition of the term \texttt{adequate}. 

\subsection{Phenomenological system}

It is never possible to measure values for all system attributes. \v{Z}ampa's approach of phenomenological system states that we do not search for any other variables than those which we can measure, otherwise we consider the system as the ``complete" system presented above. 

In our opinion, the phenomenological system is a broader issue than \v{Z}ampa considers. We demonstrate that the measurable -- phenomenological -- variables themselves include two system models: an idealized system model of a measurement and a real system model of a measuring device, cf. \cite{Stysetal2015}.

A simple illustration of the influence of a phenomenological variable point divergence gain on image processing of two consecutive micrographs obtained using widefield brightfield optical microscope from the focal plane of a live cell in order to localize interior cell objects of the size below the Abbe diffraction limit~\cite{Abbe} is shown in Appendix B.

\subsection{General system in a broad context}

It should be noted that, by the introduction of the \texttt{causal system}, \v{Z}ampa begins to constrain our set of possible abstract models. The scope of adequate models begins to be confined to those which are causal. The discussions on the origin of the arrow of time remains extensive, but \v{Z}ampa considers the causality to be one of the primitive assumptions.  

We would like to notify two important ideas about the arrow of time. One of them is the textbook principle of entropy~\cite{entropy} which states that entropy of the universe increases. Boltzmann~\cite{Boltzmann} suggested that the reason for the increase of the entropy was the fact that, in a system of large number of elements, we prevalently observe a few most frequently occurring system states and the others can be neglected. In other words, instead of examining the dynamics of each molecule, we can consider a value of state variable such as temperature or pressure as the phenomenological variable. 

Since the advent of non-linear dynamics, it is known that, even in very simple systems such as the R\"{o}ssler flow~\cite{Rossler} or the Lorenz attractor~\cite{Lorenz}, we observe a trajectory from any given set of states into one \texttt{limit set of states}, while, from the rest of the set of states, another \texttt{limit set} is reached. Such limit sets often exhibit a kind of periodic behaviour leading to a different type of ergodic state~\cite{Birkhoff} than Boltzmann expected. This state might be structured in space and observed phenomena might  be distributed in space unevenly. 

The later research in discrete dynamics~\cite{Wuensche2011,Conway_life} showed that discrete systems travel through the state space via a set of well-defined structures. Thus, \v{Z}ampa's causality principle reports only a natural fact that all dynamic systems follow a defined trajectory on their way to the limit set, where they either stop evolution or obey a \texttt{probabilistically deterministic} trajectory. Causality might be more inherent to dynamic systems than understood currently. It is a matter of ongoing research to find out whether each discrete dynamical system has a corresponding continuous system whose set of Poincar\'{e} sections is our discrete system. A trajectory of the discrete causal system is discussed in Appendix A. 

In the text below, we shall demonstrate that \v{Z}ampa's causality principle enables us to define conditions which allow to separate the system from the rest of the world. This enables us to \texttt{construct} the abstract model of the system, neither complete nor comprehensive, but always measurable. 

\section{State theory of systems}

As stated above, \v{Z}ampa's understanding of the system means to define a (complete or incomplete) set of measurable attributes, system time, and the mapping which describe the system's evolution. Attributes used by the state theory are called \texttt{state attributes} $a_i,\ i\in I$. Values, which are acquired by these attributes, are called \texttt{state variables} of the system $v_i\in V_i$, $i\in I$. The state of the system is then defined as an ordered set of $n$ values of the system state attributes. 

\subsection{System state trajectory}
In order to discriminate between causes and consequences, we differentiate two groups of state variables: The group of \texttt{inertial variables} whose attributes require time for the change of their values and the group of \texttt{non-inertial} or, not fully adequately, \texttt{static variables} whose attributes (in ideal case) do not need time for change of their values. If the set $\cal{I}$ as an ordered decomposition of the set of attributes $a_i,\ i\in I$ into sub-sets $I_i,\ i=1,2,...,m,\ m\leq n$ is in agreement with the approach in which we determine inertial and non-inertial variables, each sub-set $I_i, i=1,2,...,m$ determines either inertial or non-inertial variables.

The \texttt{system state trajectory} $z$ is defined by Equation~\ref{eq10}. The state trajectory can be decomposed into subsets $\cal{D}$ analogously as previously (Figure~\ref{fig:Causal_relations2}). According to the causality principle, for each definition set $D_{k,l}\in {\cal{D}}$, there exists exactly one definition set $C_{k,l}$ for the complete immediate cause $z\mid C_{k,l}$. 

There exists mapping $\cal{K}$,

\begin{equation}
{\cal{K}}:{\cal{D}}\rightarrow {\cal{P}}(D)
\end{equation}
which now, according to the causality law, attributes to each definition set $D_{k,l}\in{\cal{D}}$ exactly one definition set $C_{k,l}={\cal{K}}(D_{k,l})$, for which holds
\begin{enumerate}
\item
\begin{equation}
C_{k,l}\subset \bigcup_{j=1}^m D_{k-1,j},\ \ l=1,2,...,r,\ \ k=0,1,2,...,F
\end{equation}
in case of inertial variables and non-empty definition set $C_{k,l}$ and
\item   
\begin{equation}
C_{k,l}\subset \bigcup_{j=1}^{l-1} D_{k,j},\ \ l=r+1,r+2,...,m,\ \ k=0,1,2,...,F
\end{equation}
in case of non-inertial (static) variables.
\end{enumerate}

\subsection{Definition of the causal system}

The definition of the complete immediate cause is enabled by the uniqueness of demarcation of the system trajectory. The complete immediate cause can be interpreted as a model of the mechanism, according to which the system trajectory in a real system  is formed. For the abstract system ${\mathscr{S}}=(T,V)$, there is defined the mapping ${\cal{K}}$ which attributes the definition set of its complete immediate cause $C_{k,l}$ to each definition set of the consequence $D_{k,l}$ unambiguously. Such a system is called a \texttt{causal system} $\mathscr{K}$ and is identified with ordered triple $(T,V,\cal{K})$, i.e.,

\begin{equation}
\mathscr{K} = (T,V,\cal{K}).
\end{equation}

\begin{figure}[!t]
\centering
\includegraphics[width=\linewidth]{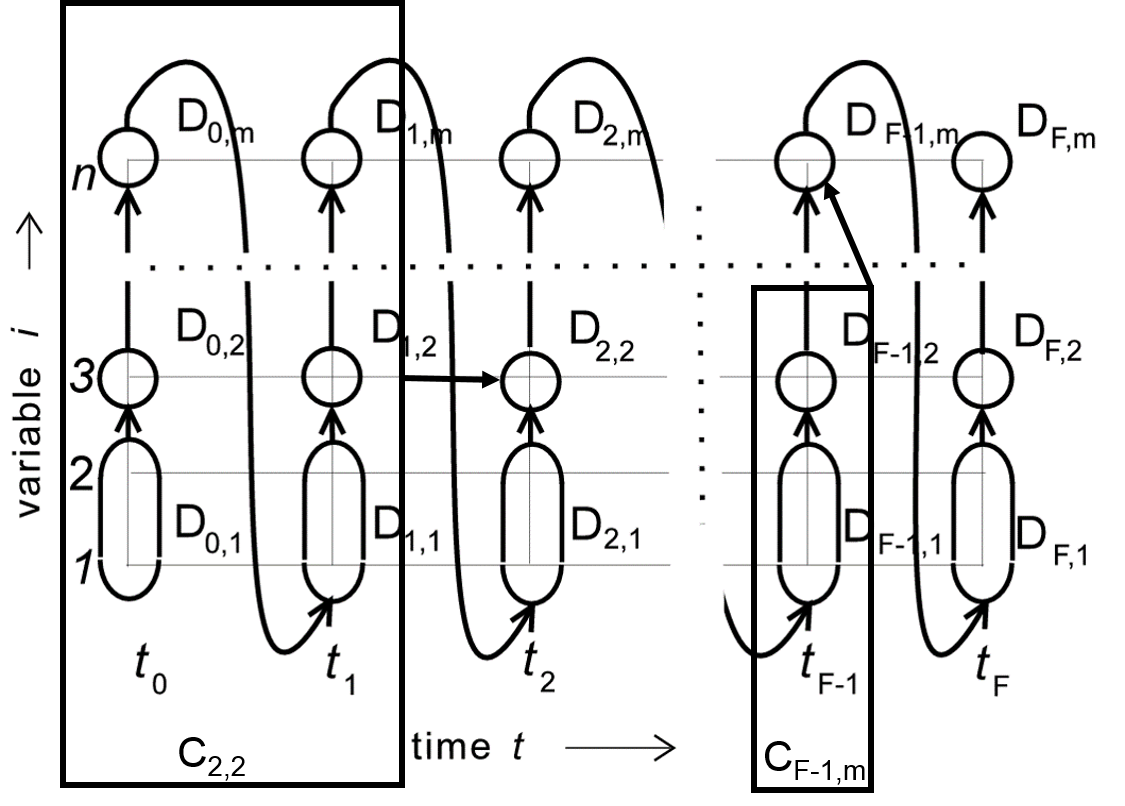}
\caption{The concept of causal relations and the importance of system model~\cite{ZampaLecture} depicted a set of variables measured at a time instant (represented by circles and ovals) and causal relation in the behaviour between measuring times. To determine the set of measured values at time $t_{l}$ we must consider not only limited a set of measured values at times $t_{j} < t_{l}$ but also an appropriate causality in intermediate times, not defined by the set of system time $T$. A set of variables at the time of measurement $D_{k,l}$ is then defined by two indexes $k,l$, where the first expresses the causality within the measurement time interval, while the other is the index of the system time from the set $I$. Then, unity $C_{k,l}\subset \bigcup_{(i,j)<(k,l)}D_{i,j}$ is \texttt{the complete immediate cause}, the set of all variables necessary for prediction of the set $D_{k,l}$. From that, among other conclusions, arises that, at least for technical reasons, the system cannot be understood without the knowledge of an appropriate system model.}
\label{fig:Causal_relations2}
\end{figure}

The ordered tuple $(C_{k,l},D_{k,l})$, $D_{k,l}\in {\cal{D}}$, is called \texttt{a system causal relation} (Figure~\ref{fig:Causal_relations2}). A set of all causal relations is then represented by mapping ${\cal{K}}$. If the cause $C_{k,l}$ is not defined for some $D_{k,l}\in {\cal{D}}$, the system ${\cal{S}}=(T,V)$ is not considered as the causal system. 

We should note that there can be cases when $z\mid D_{k,l}$ is determined by an empty mapping $z\mid \phi$, which is unique and independent of $z$, and is determined independently of any other parts of the trajectory. 
The segment $z\mid D_{0,1}$ which is called an initial condition is always defined independently.

The demarcation of the set or mapping ${\cal{K}}$ of causal relations in the system is given by the modelled reality. In certain cases, such a demarcation can be complicated. However, in some cases, the demarcation is not needed and it is enough to define an ordered decomposition ${\cal{I}}$ into $I$ as a trivial decomposition, where ${\cal{I}}=\{I\}$. Such a decomposition demarcates the trivial discrimination level of causal relations which does not enable us to model oriented relations between variables inside one time instant. This also means that, in such a system, it is not possible to define terms such as input, output, feedback bind or bond and consequently also all structural terms. Although, despite their potential incorrectness, such definitions are artificially constructed.

We show below that, in the system theory, the non-trivial decomposition of ${\cal{I}}$ of the set $I$ enables us to precisely define structural terms which can be considered as fundamental terms. 

\subsection{Complete immediate cause in a self-organizing system}

An interesting aspect of the complete immediate cause can be illustrated on a model of spatially distributed system, a stochastic cellular automaton \cite{Stysetal2016a,Stysetal2016b} described in Appendix A. This system is a typical probabilistically deterministic system $\mathscr{P}$. The evolution rule is such that the complete immediate cause $C_{k} = C_{k-1} = C_{1}$ for any $k$. There is no evolution within the time interval and the state of the system in the next time step can be completely determined from the previous time step. This description is completely true only in the final -- ergodic -- phase of the system's evolution which occurs after more than 25,000 elementary time steps. Then the actual configuration is dependent on the number and mutual positions of ignition points. In other words, the complete immediate cause $C_{k}$ covers all time steps of the simulation, where the stochastic element is added in any step as well. When the ergodic phase is reached, $C_{k} = D_{k,\nu}$, where $\nu$ is determined by the rule which governs the system.

\section{Structure of the causal system}

\emph{So far  we have considered the system as a certain enclosed and unchanged entity. In practice, it is useful to discriminate certain parts of the system and to be able to interfere in the system and change its properties. Obviously, we shall not allow any other destructive interference than those, by which the system can be decomposed into certain, from the cybernetic point of view, meaningful parts with the possibility that certain parts can be replaced by others. Later, such parts will be called sub-systems. These will be parts which are either completely isolated from the others or bound to them by special and easily disconnectable bonds transmitting only information but not mass and energy. These bonds are called information or cybernetic bonds. Parts of the system, which exchange mass and energy, are parts which are inseparable from the cybernetic point of view.}

\subsection{Bond of the causal system}

Let us have a causal system

\begin{equation}
\mathscr{K}=(T,V,{\cal{K}})
\label{eq23}
\end{equation}
(no matter whether in phenomenological or state sense) and, for all $k,\ k=0,1,2,...,F$, let us have such a $l,\ l\in \{1,2,...,m\}$ that a cause $C_{k,l}$ and a consequence $D_{k,l}$ is represented by exactly one attribute $i_l\in I$ and $j_l\in I$, respectively. 

\begin{figure}[!t]
\centering
\includegraphics[width=\linewidth]{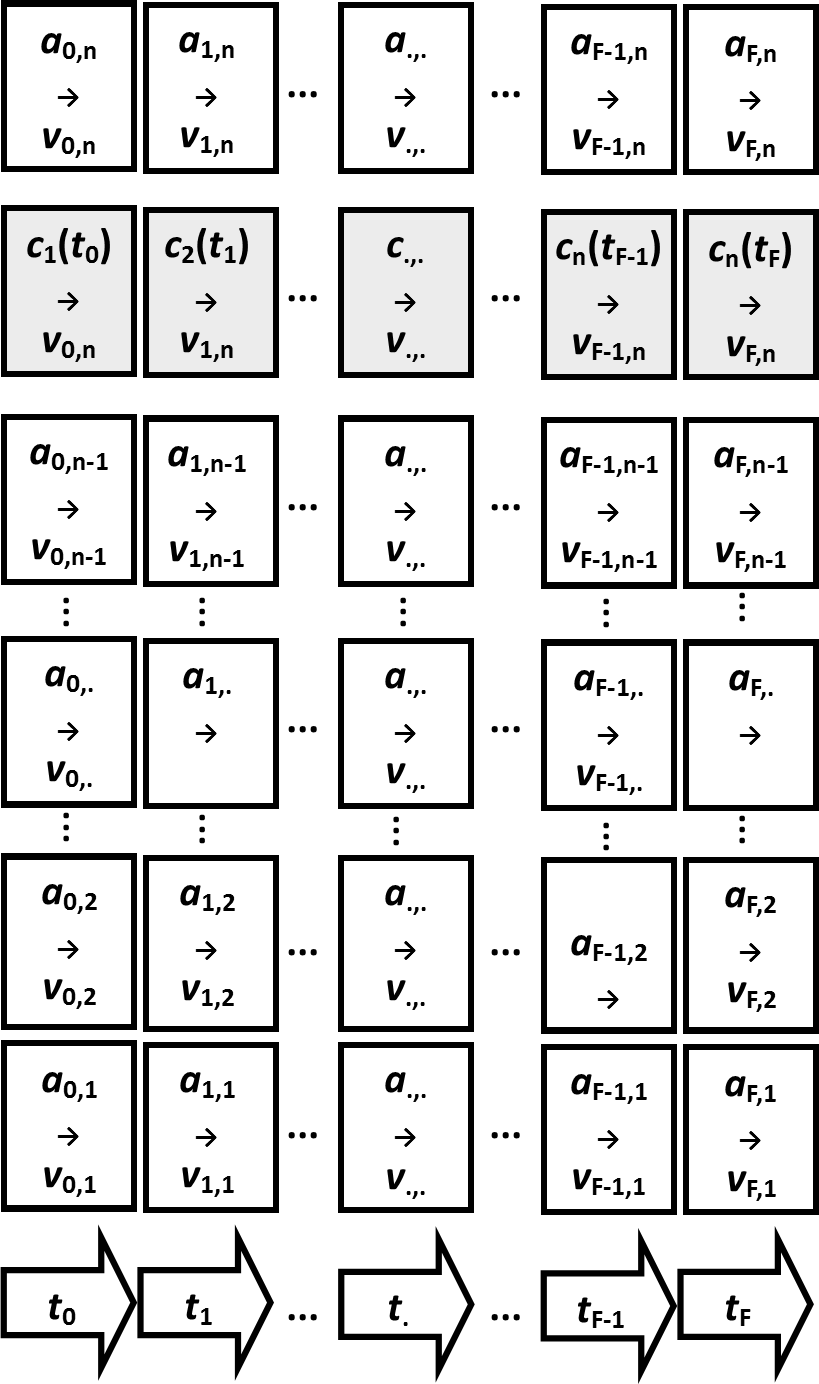}
\caption{The scheme of an (oriented) information bond $c_{.,b}$ in the system as a causal relation which is represented by exactly one attribute $a_{i,j}$ which is symbolized by one component variable $v_{i,j}$, $i \in \{0, 1, 2, ..., F\}$, $j \in \{1, 2, ..., n\}$.}
\label{fig:System_bond}
\end{figure}

Further, if the value of the attribute $i_l$ at the time $t_k$ is given by a one-component variable $v_{i_l}(t_k)\in V_{i_l}$ and the value of the attribute $j_l$ at the time $t_k$ by a one-component variable $v_{j_l}(t_k)$, then such an action by which the causal relation $(C_{k,l},D_{k,l})$ can be disconnected in the causal system $\mathscr{K}$ is assumed to be enabled. It means that, for given $k$, $l$ and consequence $D_{k,l}$, any cause $C_{k,l}$ will not exist. Then such a causal relation is called a \texttt{(information) bond of the system} $c_l(t_k)$. This information bond is oriented (Figure~\ref{fig:System_bond}).

We also assume that, at each time instant, the bond is either connected or disconnected. We should also note that it makes sense to include the attribute $a_i$ into the system only when a respective causal relation $(C_{k,l}, D_{k,l})$ is disconnectable.   

\subsection{Sub-systems of the causal system}

``In each system, parts which do not communicate mutually or which communicate only by information, not energetic, bonds can be generally found. Such parts are called \texttt{sub-systems of the causal system}." In the real world, we must only carefully classify to which extent a particular bond satisfies the requirements for the information bond. In many interesting systems from molecules up to society, such requirement can be fulfilled to a good extent. 

Let us assume a causal system $\mathscr{K}$ (Equation~\ref{eq23}) together with a set ${\cal{C}}$ of all information bonds 

\begin{equation}
c_l(t_k)\in {\cal{C}}.
\end{equation}
In this system, if we disconnect one of the sub-sets ${\cal{C}'}$ of the set ${\cal{C}}$ of all bonds, the system will split into a set of isolated parts, where each of them can become a system itself due to an appropriate choice of attributes. Namely, it will be such a sub-set of the set of all attributes of the original system whose elements lie in the given isolated part. By disconnection of each information bond, the definition set of its cause $C_{k,l}$ and thus also its cause $z\mid C_{k,l}$,  $k=0,1,2,...,F$ stop being defined for a given definition set of a consequence $D_{k,l}$. This is in contradiction with the causality principle, according to which a cause for each consequence exists. By this disconnection, the system stops being causal. According to the definition above, the new system can be considered as a sub-system which will become a causal system via complementing by an appropriate bond. It is useful to call this system a \texttt{causal sub-system}.

\texttt{A causal sub-system} is a part of a causal system which, upon the disconnection of appropriate information bonds and upon selection of appropriate primary attributes, corresponds to the basic definition of general abstract system. The causal sub-system will be marked ${\mathscr{K}}_{\iota}$, where $\iota$ is an element of an appropriate index set. The causal sub-system is often demarcated by ordered triple

\begin{equation}
{\mathscr{K}}_{\iota}=(T,V_{\iota},{\cal{K}}_{\iota}),
\end{equation}
where $V_{\iota}$ is a Cartesian product of an appropriate sub-system of a set of sub-sets $\{V_1,V_2,...,V_n\}$ and ${\cal{K}}_{\iota}$ is an appropriate sub-set of the set ${\cal{K}}$ of all causal relations of the original causal system $\mathscr{K}$.

It is important to note that the steps described above serve only for a decomposition of the system into sub-systems. A system is always causal and, therefore, also physically possible. A non-causal situation leads to a result without a cause. This situation occurs only in one part of the system trajectory -- in the initial state, at the beginning of the state trajectory. Here, we distinguish two kinds of system's behaviour -- static and dynamic. To show that the system is static, it is necessary to have at least two identical consecutive states. The different situation is in case of the dynamical systems, where, for the definition of the initial state, it is necessary to define the initial conditions with a cause resulting from the initial state. However, the initial conditions are given -- they are ad hoc results -- without any cause, even if there were causal relations which would lead to the initial conditions in a given way. 

\subsection{Inputs, outputs and internal attributes}

An attribute of a causal sub-system, which is a consequence of a certain system bond, is called an \texttt{input} of this sub-system and a variable, which represents a value of the input at each time instant $t_k\in T$, is called the \texttt{input variable} of the sub-system $u(t_k)$. An attribute of the causal system, which is a cause of a certain bond, will be called an \texttt{output} of a given sub-system and a variable, which represents a value of the output at each time instant $t_k\in T$, will be called an \texttt{output variable} of the sub-system $y(t_k)$. A general attribute is not necessarily an input or output of a causal sub-system. An attribute, which is neither input nor output of a sub-system, is called an \texttt{internal} attribute of the sub-system and the corresponding variable is called an \texttt{internal variable} of the sub-system $x(t_k)$.

Since all variables are defined in the causal system, each input must be connected to one of the outputs. From that it is clear that the causal system (not a sub-system) has as many bonds as inputs. From the definition of outputs also comes out that any (as well as zero) number of inputs can be connected to each output. 

The precise definition of the \texttt{system bond} and the \texttt{information bond} is the key concept. In order to construct an abstract dynamic system, we can easily make a mistake in examining the system which is inseparable from other parts of the physical nature. Any measurement in such a system is obviously impossible.

\subsection{Measurable sub-systems}

The concept of decomposition of the system into sub-systems was developed for the purpose of the system control. However, in the system control, the errors introduced by omission of the properties of a control device can be often overlooked. The decomposition into sub-systems is much more critical in the case of measurement.

In Appendix C, we show an analysis of the signal of the HPLC-MS device as a prominent example of the usage of the device in the chemical analysis.  


\section{Conclusions}

If we intend to use mathematics for the description of dynamic systems, the best known guidance is qualitative dynamics~\cite{Poincare1880,chaosbook}. Qualitative dynamics teaches us that a phase space is sectioned into zones of attraction of different limit sets. It tells us, which types of behaviour we can anticipate, if we wait sufficiently long, until a system arrives at its limit set and stays there. The quality of the limit set is ergodic behaviour~\cite{Birkhoff}, i.e. that, a during sufficiently long time, the system visits each of its states at least once. In layman's terms, the system resides in a stable, oscillating, or other repeatedly changing state. 

It seems that most systems are on the trajectory towards the limit set~\cite{Stysetal2015}. Qualitative analysis of limit sets can be transformed into discrete problem by sectioning of the trajectory by Poincar\'{e} sections. They can be equally well used for sectioning the trajectories which lead to the limit sets. But these trajectories are seldom analyzed. Thus, it is not clear whether the timely structured states and "confluences" of trajectories, which originate from different gardens of Eden in discrete dynamics~\cite{Wuensche2011}, do always have their continuous dynamic counterparts related by Poincar\'{e} sectioning. If it was so, our freedom in choosing a model would be constrained severely~\cite{Stysetal2015b}. The only question which remains is whether the Nature is constrained by mathematics. 

It is certain that \v{Z}ampa himself focused his thoughts on technical systems. He never searched for limit sets and did not discuss the qualitative dynamics. Instead, \v{Z}ampa examined limits for the existence of the causal system and conditions by which the system can be separated from the rest of the universe. Thus, \v{Z}ampa's state theory complements our scope of objective constraints for the choice of models. 

\v{Z}ampa's state theory is also very close to measurement. The phenomenological system addresses the problem directly, but it is certain that an adequate abstract system cannot be constructed only using variables which we -- by chance -- measure. By accepting this, we can construct a machine which performs what we engineered. But, to our surprise, the machine will fail so often and the final marketed construction will include many aspects which the first constructor did not anticipate. Also, the same holds for many elements which the constructor did not even notice and are hidden inside his equations and software which he used. The chief constructor can remain confident that his model is still the core of his machine, but the outside reality, following the rules of qualitative dynamics outside the limit set, will continue by its own way. 

May we know how to proceed? The knowledge of qualitative dynamics, as Predrag Cvitanovi\v{c} expressed, has "holes large enough to steam a Eurostar train through them"~\cite{chaosbook}. Qualitative dynamics concerns with limit sets and seldom studies the trajectory through the zone of attraction towards them. The exceptions are Wuensche's 8-level discrete dynamic networks and multilevel systems~\cite{Stysetal2015b}. As proved~\cite{Stysetal2015b}, the scope of types of qualitatively different trajectories is limited and much smaller than usually considered for potential technical constructions. 

\v{Z}ampa gives a part of the recipe. First, we must be able to dissect the system into the sub-systems which are mutually bound only by an information bond. This gives us the structure of the system for studying. A human in the society or a bird in a flock can survive independently of the context, at least during the time of the measurement. Its scope of trajectories can be studied. In societal context, when the information bond(s) are connected, trajectories which in disconnected systems are sparse or non-existent can suddenly prevail. 

Second, after the dissection of the system into the sub-systems, we must look for a system trajectory. For each set of values of a particular variable, we must determine the complete immediate cause. In traditional dynamics, for a system of the unknown technical composition, this is a neverending process. Coming from technical environment, \v{Z}ampa considers knowledge of the technical limit of the range of attribute states. The first advice for examination of the systems which we do not know is to search for the complete immediate cause. Any relevant model must consider an adequate structure of inertial and non-inertial variables such that each trajectory's segment has its own complete immediate cause. Unfortunately, in case of, e.g., many biological or societal systems, the complete immediate cause precedes the start of our experiment or observation~\cite{Stysetal2015}.

\section{Appendices}

System description is not only useful for artificial constructed systems, but also for the data measured in the experiments, especially, if they are of biological or chemical origin. In Appendices, we introduce three illustrative examples of data description, which help biophysical understanding of the measured natural phenomena. The first example is the famous dynamic Belousov-Zhabotinsky reaction, which is observed in the discrete time and where variables derived from the R\'{e}nyi entropies are considered as state variables describing the observed states. The evolution of the system is thus reflected in changes of information of the underlying physico-chemical phenomena. The next model considers microscopy as an information channel, where time-spatial evolution of the point spread function of a live biological specimen is approximated by an image z-stack of the light distribution in the in-focus region. The last system is a model of liquid chromatography in tandem with mass spectrometry, where the retention time of the chromatographic elution represents the evolution of the compound composition in the dependency on the set-up of the chromatograph. The additional conditional segmentation of the dataset into sub-systems follows the system approach and represents relevant chemical decomposition of the measurement. In all cases, the adopted system approach leads to the unscrambling of the system. The demonstrated system decompositions can provide meaningful information about the measured experiments.

\subsection{Appendix A: The role of the system time: the analysis of a model series of the Belousov-Zhabotinsky reaction}

The Belousov-Zhabotinsky reaction is a prominent experimental example of the self-organization in the nature~\cite{Zhabotinsky1959,ZhabotinskyRovinsky}. One of its possible models is a noisy hodgepodge machine~\cite{Stysetal2016a,Stysetal2016b} which can simulate its complete trajectory. This type of simulation provides an image series with the shortest possible time interval between two consecutive images.

\begin{figure*}
\centering
\includegraphics[width=\linewidth]{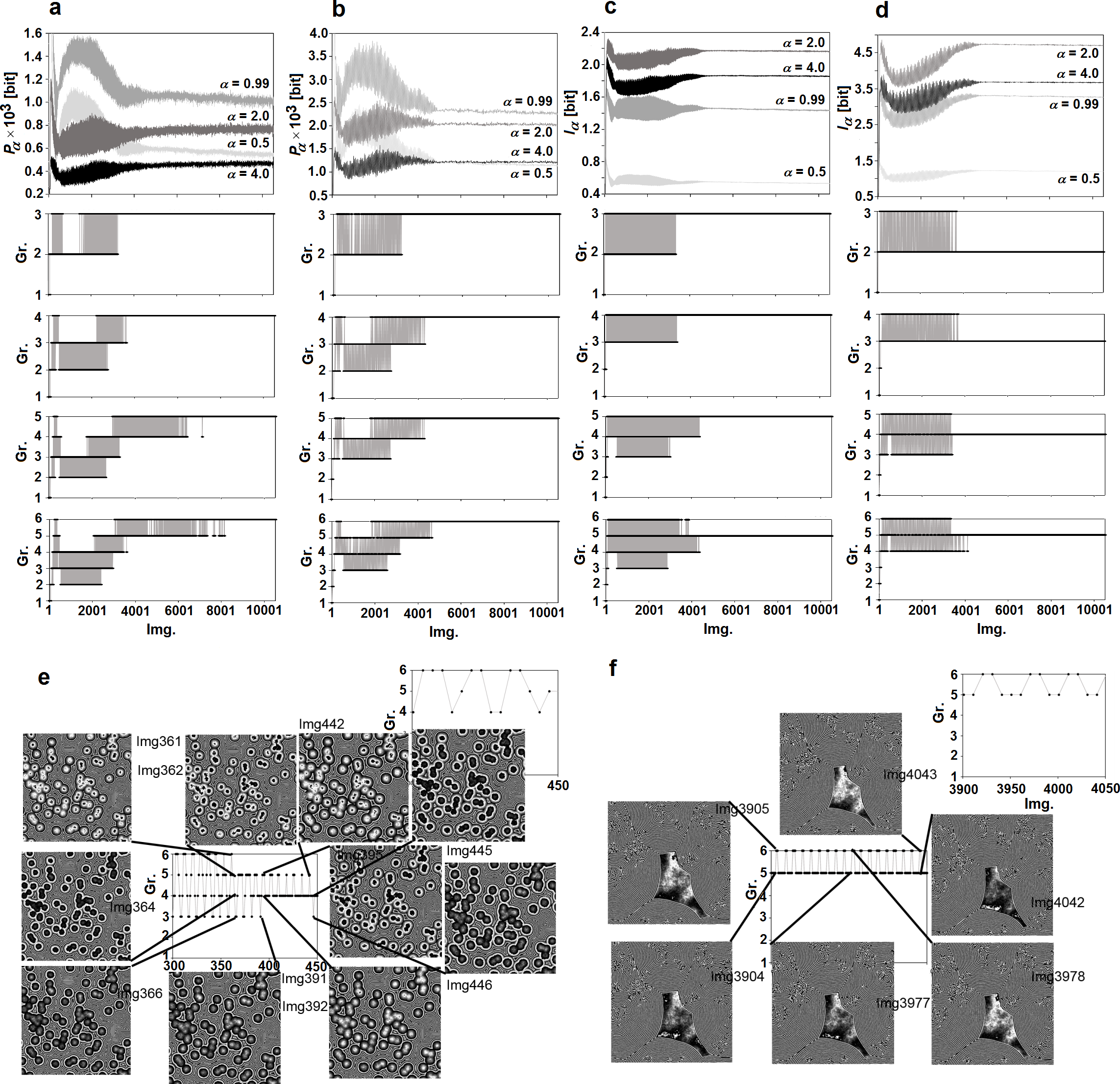}
\caption{k-Means clustering (squared Euclidian distance) of an image series of a stochastic hodgepodge model of the Belousov-Zhabotinsky reaction according to similar structures using a point divergence gain entropy density $P_{\alpha}$ (\textbf{a--b}) and point divergence gain entropy $I_{\alpha}$ (\textbf{c--d})  for $\alpha$ = \{0.1, 0.3, 0.5, 0.7, 0.99, 1.3, 1.5, 1.7, 2.0, 2.5, 3.0, 3.5, 4.0\}.  Analysis of a full-image series\textbf{a,c}) and of every 10$^{\mbox{th}}$ image of the full-image series \textbf{b,d}). \textbf{Row 1a--d}) Examples of time courses of $P_{\alpha}$ and $I_{\alpha}$, respectively, for $\alpha$ = \{0.99, 2.0, 0.5, 4.0\}. \textbf{Row 2--5, a--d}) Clustering of images into 3--6 groups, respectively. \textbf{e--f}) Sections of the trajectories shown in \textbf{Row 5c--d}. No. of image corresponds to the time unit.}
\label{fig:PDGE_series_BZ_model} 
\end{figure*}

For analysis of the image series, we have developed a method of the calculation of the point divergence gain entropy ($I_\alpha$) and the point divergence gain entropy density ($P_\alpha$) which are cumulative variables derived from the absolute values of the point divergence gain ($\omega_{\alpha,x,y}$)~\cite{Rychtarikova2018,Rychtarikova2015,Rychtarikova2016}. The parameter-dependent spectra of these macroscopic variables characterize stepwise differences in image series and evaluate the evolution of the system between two consecutive data points. In Figure~\ref{fig:PDGE_series_BZ_model}, the set of parameter $\alpha$ contained values 0.1, 0.3, 0.5, 0.7, 0.99, 1.3, 1.5, 1.7, 2.0, 2.5, 3.0, 3.5, 4.0. Next, the image series, where each image is described by the parameter-dependent spectrum, is characterized by cluster analysis (k-means with squared Euclidian distance) which separates the series into groups of similar images. Then, each group represents the section of the trajectory (Figure~\ref{fig:Causal_relations}). 

The statistical analysis in Figure~\ref{fig:PDGE_series_BZ_model} compares results for $I_\alpha$ and $P_\alpha$ which are computed from full-image series and from series, where the divergences were calculated for every 10$^{\mbox{th}}$ image. The analysis of the full dataset is typical of oscillations between clusters (Figure~\ref{fig:PDGE_series_BZ_model}\textbf{a,c}). The detailed inspection (Figure \ref{fig:PDGE_series_BZ_model}\textbf{e--f}) found that these oscillations are due to the dominant color of structures which surround the ignition point. These changes are slower than the frequencies of oscillations of dense square waves which dominate the early phase (Figure~\ref{fig:PDGE_series_BZ_model}\textbf{e}) and of the circular waves which dominate the late phase (Figure~\ref{fig:PDGE_series_BZ_model}\textbf{f}). In the decimated trajectory, this feature is unrecognizable, although the change of the prevalent structure is determined properly.

The differences between the original and the decimated series calculated from $P_\alpha$ (Figure~\ref{fig:PDGE_series_BZ_model}\textbf{a--b}) are more pronounced than in the case of $I_{\alpha}$ (Figure~\ref{fig:PDGE_series_BZ_model}\textbf{c--d}). This is a consequence of the fact that, in the $I_{\alpha}$ calculation, we sum the values $\omega_{\alpha,x,y}$ for all points (pixels) and the result is dominated by frequent values whose occurrences do not change significantly over 10 consecutive images. The $P_{\alpha}$ sums only different levels and, therefore, is sensitive to subtle changes which give rise to new unique $\omega_{\alpha,x,y}$ values. For higher number of clusters, the results of the clustering vectors $I_\alpha$ and $P_\alpha$ are dramatically different.


Despite the complete and the decimated trajectories are similar, at first sight, the sole usage of the sparse system time $T$ leads to the completely different structure of the trajectory. It demonstrates that the choice of the system time is not voluntary, but, in many cases, the time have to be measured at elementary system events. Simple rules such as the Shannon-Nyquist Sampling Theorem give a good advice for that. Nevertheless, as the given example shows, even if we detect the high-frequency changes of variables, there can be low-frequency changes of variables which might not be fully understood and which significantly change the result of the analysis.  

From many real datasets, where the sampling frequency is technically limited, the model of the system can be never deduced in the straightforward manner. This exactly illustrates the idea of the \texttt{complete immediate cause} $D_{k,l}$ (Figure~\ref{fig:Causal_relations}). For the prediction of the succeeding system's behaviour, it is necessary to know not only the system's state at a few preceding time instants, but also a good system's model which predicts the system's behaviour between the measured time points $l$, i.e., between the values of the system time, at the time instants $k$. In the particular case of the model of the Belousov-Zhabotinsky reaction, which anticipates time and spatial discreteness, we know the probability distribution function of the time element. The question to which extent the existence of the element of time unit is general remains to be answered by the fundamental research of physics.  
 
\subsection{Appendix B: Phenomenological variables unravel superresolved structures in a series of standard brightfield micrographs of a live cell}


\begin{figure*}
\centering
\includegraphics[width=\linewidth]{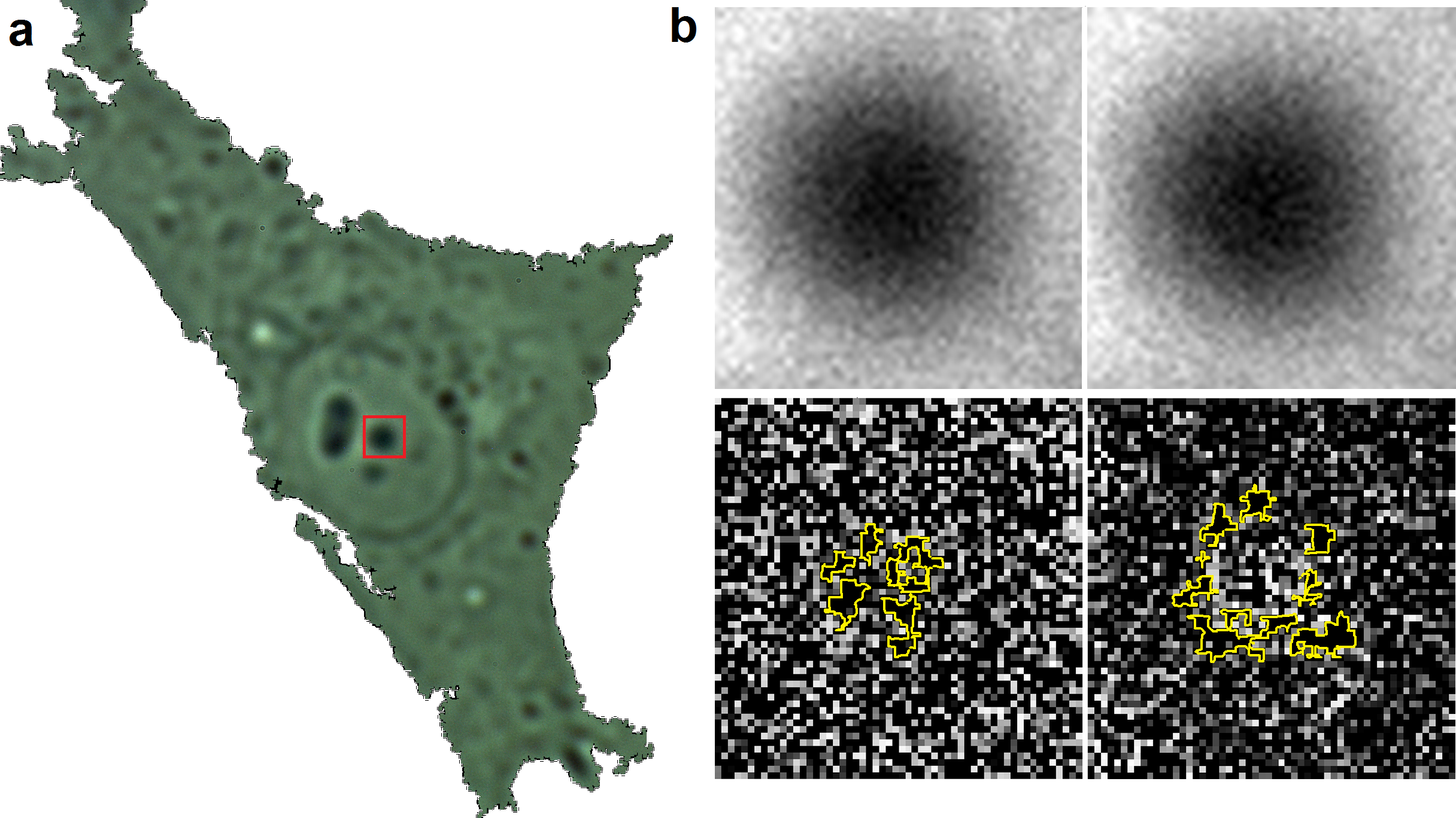}
\caption{Superresolved localization of the internal structures of a living MG-63 cell's nucleolus. \textbf{a}) The original image of a living MG-63 cell in the focal region of a brightfield microscope. The image was captured by a colour digital camera as a 12-bit raw matrix. For visual inspection, the 12-bit image was transformed into an 8-bit image by the Least Information Lost (LIL) algorithm \cite{Nahlik2016}. The red square highlights the section of nucleolus used for calculation of $\omega_{0.99,x,y}$ values. \textbf{b}) $\omega_{0.99,x,y}$-Transformation of intensities in the red camera channel of two consecutive images in the cell's focal region. \textbf{Upper row} -- Original intensities from which the $\omega_{0.99,x,y}$ values (\textbf{lower row}) were calculated. \textbf{Lower row} -- Negative (\textbf{left}) and positive (\textbf{right}) values of $\omega_{0.99,x,y}$. The yellow demarcated regions corresponds to the structures which are attributed to be chromatin by electron microscopy. Scaling of the original 16-bit intensity image depicting $\omega_{0.99,x,y}$ into an 8-bit computer screen by the simple sectioning of the scale (\textbf{lower left}) and by the LIL transformation with the fully utilized 8-bit scale (\textbf{lower right}). The brightest and darkest intensities correspond to the $\omega_{0.99,x,y}$ values for the time-stable and unstable structures, resp. The size of the image voxel is 64 $\times$ 64 $\times$ 130 nm$^3$.}
\label{fig:38_01_lil}
\end{figure*}

The only procedure, which we can use for the description of the cell dynamics, is an extensive analysis of the image information in the brightfield micrographs. The brightfield optical microscope transmits light, modifies the wavefront, and the intensity of the electromagnetic field (i.e., the probability amplitude of the occurrence of photons at a given position in the space) is captured by a detector as a near-continuous signal. The number of resulted charge transfers is detected and transformed by an analog-digital converter into a matrix of numbers. Therefore, we need to analyze a phenomenological discrete variable (color of an image pixel) instead of a near-continuous variable (number of photons in space).

Aspects of capturing an image by a digital camera, which is a sub-system on its own, are discussed in~\cite{Nahlik2016}. The observed intensities represent a sum of changes of the intensities which pass a sample and a optical path of the microscope. An actual shape of the observed object is determined by diffraction properties and by the deformation of the electromagnetic field along the optical paths. According to the Mie theory~\cite{Mie1909}, a larger object does not always scatter more light and its scattered image is not always larger. The dependency of the scattering efficiency on particle's size has several maxima. The Nijboer-Zernike theory~\cite{NijboerZernike1944} tries to explain the distortion of the wavefront by its traveling along the optical path. In fact, the description of the behaviour of a complicated -- patterned -- wavefront is not satisfactory. We do not have any theory which would describe the resulted pattern sufficiently, not even for a very simple diffracting object. Thus, the observed object together with the microscope represent a system which has to be examined from \v{Z}ampa's point of view.

However, since both the diffraction and the passage of light through the optical system of the microscope are changes of the electromagnetic field, they can be hardly disconnected into sub-systems. In addition, it was found that it is possible to detect objects far below the diffraction limit of light and the diffraction limit decreases with increasing light intensity~\cite{Robertson1969}. The latter observation is difficult to explain by any existent theory.

Articles~\cite{Rychtarikova2015,Rychtarikova2016} describe a method how to obtain superresolved micrographs, i.e., locations of objects, with the precision below 50 nm, from a series of the most ordinary brightfield widefield optical microscopic images. The series was obtained by movement of the biological specimen along optical axis of the microscope. As similar to Appendix A, the change of information (intensity) between two consecutive images  is for each camera pixel analyzed using the point divergence gain. Wherever the information remains unchanged, we assume that we have localized an object whose response is larger than a voxel, i.e., than the area of the camera pixel multiplied by the z-step. 
The intensity change between two consecutive z-stack images of a live cell has basically twofold origin: (1) the diffraction response of the observed objects is thinner than the z-step or (2) the object moved during the image capture. Our next assumption for the choice of a macroscopic phenomenological variable was that the image structure is multifractal. The chosen method of analysis~\cite{Rychtarikova2015} is further based on an assumption that two image points of identical intensity lying at two levels directly above each other represent (with a high probability) the same information. The $\alpha$ of different values used in the calculation result in quite different histograms of ($\omega_{\alpha,x,y}$): a low $\alpha$ separates rare points, while a high $\alpha$ separates more frequent points. The lower row of Figure~\ref{fig:38_01_lil}\textbf{b} illustrates values of the point divergence gain at $\alpha = 0.99$ ($\omega_{0.99,x,y}$) which correspond to the information difference due to the replacement of a pixel's intensity in the first image by the pixel's intensity at the same position in the second image for a couple of images in the focal region differing from each other by the z-step of 130 nm (the higher the $\omega_{0.99,x,y}$, the brighter the pixel). The most important result of this image recalculation is the highlighting a fine structure inside a nucleolus which is similar to granular and fibrilar structures reported by canonical electron microscopy~\cite{image}. This fine structure is not visible in a 8-bit color representation but in a 12-bit intensity file. This shows that, inside the image of the organelle, there are different values of $\omega_{0.99,x,y}$ which bring information about different diffracting objects. It confirms a finding~\cite{Robertson1969} that, in a brightfield optical micrograph, objects of the diameter of 25 nm is discriminable~\cite{Urban2014}.

The analysis described above indicates that any information about a live cell provided by the microscope is a highly phenomenological quantity which reflects the content of chemical compounds at the particular point and at the neighboring points. In other words, the transformation of the color-coded image series into $\omega_{\alpha,x,y}$-coded images provides a set of measurable (phenomenological) variables which maximally yields the information brought by the biological experiment.  


A visual inspection of an uncalibrated brightfield micrograph  and the phenomenological character of the image information-entropic variables also have a purely technical reason (see Fig \ref{fig:38_01_lil}\textbf{a}). Vice-bit images are typically stored and visualized in a 8-bit format. As described in \cite{Nahlik2016}, even in a lossless compression, a series images are transformed by an algorithm which differs for each image. The alternative -- a simple sectioning of original (12- or 16-bit) image into 256 levels -- can leave most levels unoccupied. It allows us to conclude that a majority of images analyzed in the world is transformed in an uncontrolled way and, therefore, useless for an exact analysis.

\subsection{Appendix C: Information bond in a dataset obtained from liquid chromatography with mass spectrometry (HPLC-MS)}

\begin{figure}[!t]
\centering
\includegraphics[width=0.85\linewidth]{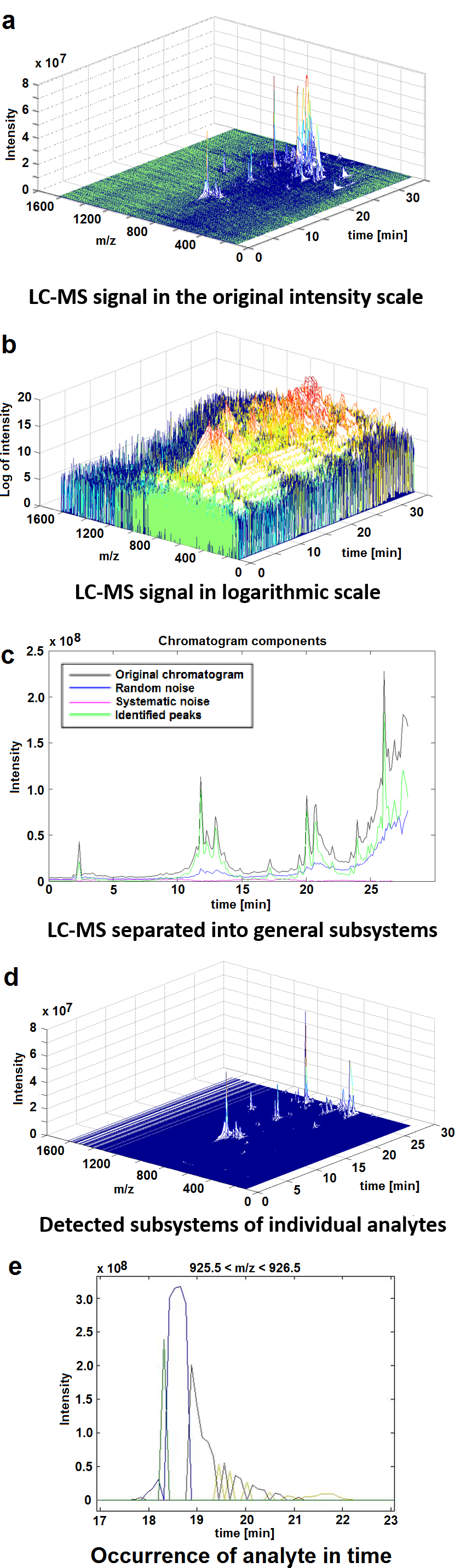}
\caption{Signal processing in HPLC-MS.}
\label{fig:LC-MS_composite_vertical}
\end{figure}

In HPLC-MS analysis, we consider an information obtained from a measured signal as a system~\cite{Urbanetal2009} which can be separated into sub-systems. The LC-MS measurement~\cite{LC-MStextbook} is a combination of a physico-chemical chromatographic separation of individual compounds on a chromatographic column, a separation of ionized molecules upon flight through the electromagnetic field, and a detection of the number of molecules by a detector. There are numerous technical realizations of this experiment. The analysis which is presented here is general and is not dependent on a concrete technical set-up. The resulted dataset is a two-dimensional set of values (Figure~\ref{fig:LC-MS_composite_vertical}\textbf{a--b}).

The key idea is that the detected signal is a set of system attributes $a_i, i\in I$, where $I$ is a set of measured time instants. At each time instant $t$, we obtain one intensity value $y(m,t)$ for each value of mass $m$. Each point $(m,t)$ is assigned at least into one of these sub-systems: 

\begin{enumerate}  
\item random chemical noise $r(m,t)$,
\item systematic noise $q(m,t)$,
\item signal $s(m,t)$. 
\end{enumerate}

The random chemical noise $r(m,t)$ is prevalently a combination of chemical compounds randomly eluted from the column with electrical noise on the detector. This kind of noise can be modelled by a standard distributions, e.g. by a log-normal distribution as in the given case (Figure~\ref{fig:LC-MS_composite_vertical}). The information about the random chemical noise is contained in ca. 93\% of data points. It means that the properties of free random chemical noise can be determined with a high reliability.

The systematic noise $q(m,t)$ and the signal $s(m,t)$ are formed by systematic responses of specific molecules which reach the detector. The systematic noise $q(m,t)$ represents molecules which are constantly present in the solvent. This noise occurs as forms of ridges at the positions of specific \textit{m/z} (molecular mass to charge) ratios as well as of chromatographic peaks in each empty run. The ridges can be clearly assigned to the systematic noise and form the signal component $q(m,t)$.

The signal $s(m,t)$ is the sought response, i.e., the signal originating from molecules (analytes) which were added to the chromatographic apparatus and should be detected. The signal of the analyte appears as a chromatographic peak of a specific \textit{m/z} value. For the purpose of the analysis, despite the qualitatively different origin, the chromatographic peaks of both the background and the analyte are summed in the signal $s(m,t)$. 

The signal components $r(m,t)$, $q(m,t)$, and $s(m,t)$ are not independent. Namely, we observe a decrease of the $q(m,t)$ and an increase of the $r(m,t)$ at the position of the non-zero $s(m,t)$ (Figure~\ref{fig:LC-MS_composite_vertical}\textbf{c}). Moreover, the $s(m,t)$ can be further separated into signals of individual molecules which appear as a few parallel peaks of different adducts, isotopologues, etc. 

Since the $r(m,t)$, the $q(m,t)$, and components of the $s(m,t)$ represent sub-systems, the signal $y(m,t)$ is a sum $y(m,t) = r(m,t) + q(m,t) + s(m,t)$. It should be noted that the mapping $y:T \times M \rightarrow I$ is probabilistic. Each contributing component, including the responses of individual compounds, shows a probability density function which is not generally known. 

The first step of the signal analysis is to find the envelope of the distribution function of the $r(m,t)$ which can be determined with a high level of confidence. Any signal outside this envelope belongs to either the $q(m,t)$, the $s(m,t)$ or an isolated electrical spike. These spikes are confined to one point, they are not surrounded by other signals differing from the $r(m,t)$ and are easily eliminated. The noise $q(m,t)$ appears for almost all $t$ values, while the $s(m,t)$ appears only timely. For both cases, an algorithm had to be found in order to analyze the features of the signal. Figure~\ref{fig:LC-MS_composite_vertical}\textbf{e} shows an example of a course of a probability peak $s(m,t)$ after subtraction of the $r(m,t)$.

Despite the fact that the $r(m,t)$, the $q(m,t)$, and the $s(m,t)$ are strongly interconnected and affect each other, the analysis using the assumption of observation of the probabilistic system which can be separated into sub-systems was successful. Figure~\ref{fig:LC-MS_composite_vertical}\textbf{d} depicts that each peak of analyte was separated and many signals originally hidden in the background noise were found. Moreover, compounds identified in the control experiments (i.e., empty run or blank) were identified, used for calibration, and removed from the final set of analytes' signals. The true information bond is never practically realized in physical systems. Nevertheless, it can be often assumed in the analysis of a real dataset. This explains the success of system analyses of systems with an input and an output in which the properties of the input and the output are not considered.

\begin{nomenclature}
\item{a_i}{Abstract attribute, e.g., a coordinate of position, a coordinate of speed, verity of statement}
\item{A}{Set of all abstract attributes}
\item{B}{Sub-set of the set of all systems trajectories $\Omega$}
\item{\cal{B}}{Set of all system events}
\item{c_l(t_k)}{Information bond of the system}
\item{\cal{C}}{Set of all information bonds}
\item{C_{k,l}}{Cause}
\item{D}{Definition set of a mapping of a system trajectory $z$}
\item{D_{k,l}}{Consequence}
\item{\cal{D}}{Ordered decomposition of the set $D$}
\item{\mathscr{D}}{Deterministic abstract system}
\item{I}{Set of indexes of attributes}
\item{I_{\alpha}}{Point divergence gain entropy}
\item{\cal{I}}{Set defined above $I_j$, $j = 1, 2,..., m$ with a causal condition <}
\item{\cal{K}}{Causal mapping}
\item{\mathscr{K}}{Causal system}
\item{m}{Molecular weight}
\item{m/z}{Mass-to-charge ration}
\item{P(B)}{Probability of each event B in the set \cal{B}}
\item{\mathscr{P}}{Stochastic abstract system}
\item{q(m,t)}{Systematic noise}
\item{r(m,t)}{Random chemical noise}
\item{s(m,t)}{Signal}
\item{\mathscr{S}}{Abstract system}
\item{t}{Real time}
\item{T}{$K$-element set of real times with indexes 0, 1,..., $F$}
\item{u(t_k)}{Input variable of the sub-system}
\item{v}{System variable}
\item{V}{Definition set of system variables}
\item{v_i}{Abstract variable}
\item{V_i}{Definition set of abstract variables}
\item{y(m,t)}{Additive signal}
\item{x(t_k)}{Internal variable of the sub-system}
\item{y(t_k)}{Output variable of the sub-system}
\item{z}{System state trajectory}
\item{\alpha}{R\'{e}nyi coefficient}
\item{\iota}{Element of an appropriate index set for a causal sub-system}
\item{\nu}{Order of the consequence in the ergodic phase}
\item{\phi}{Empty set}
\item{\omega_{\alpha,x,y}}{Point divergence gain}
\item{P_{\alpha}}{Point divergence gain entropy density}
\item{\Omega}{Set of all system trajectories}
\end{nomenclature}

\begin{acknowledgements}
This work was supported by the Ministry of Education, Youth and Sports of the Czech Republic---projects CENAKVA (No. CZ.1.05/2.1.00/01.0024), CENAKVA II (No. LO1205 under the {NPU} I program), the CENAKVA Centre
Development (No. CZ.1.05/2.1.00/19.0380)---and from the European Regional Development Fund in frame of the project Kompetenzzentrum MechanoBiologie (ATCZ133) in the Interreg V-A Austria---Czech Republic programme.
\end{acknowledgements}


\end{document}